\documentclass[aps,pra,nofootinbib,preprint,amsmath,amssymb,floatfix]{revtex4}
\usepackage{color}
\usepackage{graphicx}

\begin{document}

\newcommand{\tc}{\textcolor}
\newcommand{\g}{blue}
\newcommand{\ve}{\varepsilon}
\title{Casimir force between ideal metal plates in a chiral vacuum}         

\author{Johan S. H\o ye$^1$ and  Iver Brevik$^2$  }      
\affiliation{$^1$Department of Physics, Norwegian University of Science and Technology, N-7491 Trondheim, Norway}

\affiliation{$^2$Department of Energy and Process Engineering, Norwegian University of Science and Technology, N-7491 Trondheim, Norway}
\date{\today}          

\begin{abstract}
We calculate the Casimir force between two parallel ideal metal plates when there is an intervening chiral medium present. Making use of methods of quantum statistical mechanics we show how the force can be found in a simple and compact way. The expression for the force is in agreement with that obtained recently by Q.-D. Jiang and F. Wilczek [Phys. Rev. B {\bf 99}, 125403 (2019)], in their case with the use of Green function methods.
\end{abstract}
\maketitle

\bigskip
\section{Introduction}
\label{secintro}

Chiral materials attract at present considerable interest. They are characterized by right-circular polarized photons and left-circular polarized photons behaving differently. Optically active materials found in nature are examples of chiral materials, and the Faraday rod in which a longitudinal magnetic field is the essential ingredient, enabling the photon polarization plane to change as the pulse propagates, is well known.

The topic to be studied in this paper, is how the Casimir effect between two parallel metal plates becomes  influenced by the intervening space (medium, or for simplicity just a  vacuum) when this space has chirality properties. The Casimir force; its strength as well as its sign, will turn out to dependent on the optically active components in the permittivity tensor, the \tc{red}{width} of the chiral gap, and also the strength of the longitudinal magnetic field in the case of a Faraday medium. We will consider  arbitrary temperatures.   This problem was recently analyzed by Jiang and Wilczek \cite{jiang19}. By applying Green function methods for scalar fields as well as for electromagnetic fields, {they} showed how the Casimir force for {a} Faraday medium  depends on the parameters mentioned above, namely the gap {width} and the magnetic field magnitude via the Verdet constant.

Our main motivation for considering the chirality Casimir problem anew, comes from the observation that the situation is similar to that of Casimir interaction between a dielectric and a magnetic thick plate. This is the classic problem, often called the Boyer problem \cite{boyer74}, where the force is known to be repulsive.  In a recent work we did a detailed evaluation of the force in the Boyer configuration, finding agreement  with the  Lifshitz formula for this situation \cite{hoye18}. This was done by use of the   statistical mechanical method  where the quantum problem via the Feynman path integral becomes equivalent to a classical polymer problem. By that a quantized harmonic oscillator reduces to a set of classical harmonic oscillators, one for each Matsubara frequency \cite{hoye81}. As before, since the electromagnetic field is linear, it can be eliminated to be replaced with (frequency dependent) dipolar interactions between polarizable particles. As usual, to obtain induced interactions between separate media, the particle structure can be replaced by the dielectric constants of the continua. This statistical mechanical way of approaching the Casimir problem turns out in general to  be  compact and effective, and has earlier been made use of by the present authors under various occasions \cite{hoye13,hoye92,hoye93,brevik88,hoye98}.

Consider now the conventional Boyer configuration, where there is one dielectric plate and one magnetic plate (both plates thick) separated by a vacuum gap. By radiation from the constituent dipoles, the magnetic field is transverse to the electric field. Imagine that the gap is instead filled with a chiral medium arranged such that the plane of polarization of the photons is rotated by just $90\,^o$. Then the situation becomes similar to that for the Boyer case, and implies that the magnetic plate can be replaced by an equivalent dielectric plate. By rotating the latter dielectric plate  $90\,^o$,  the usual dielectric interaction between a pair of electric dipoles is recovered. In order to determine the {resulting} interaction, one moreover has to figure out what happens by radiation in the opposite direction. If the polarization plane rotates $90\,^o$ back to its original direction, the standard Casimir force will not be influenced. If, by contrast, the interaction {continues}  to add another  $90\,^o$ rotation to obtain a $180\,^o$ rotation, the Casimir force will be modified to be repulsive. This is the Boyer repulsive effect {for a dielectric and a magnetic plate.}

This situation an be generalized, in the sense that one can imagine rotation angles of the polarization plane different from $90\,^o$. For arbitrary rotation angles, and for ideal metal surfaces, the {magnetic model of Ref.~\cite{hoye18}} can be extended in a rather straightforward way. In this case  the TM (transverse magnetic) and TE (transverse electric) modes between the plates are equivalent as they both have equal reflection coefficients (equal to one) for non-zero frequencies. (For zero frequency only the electrostatic TM mode is present.)

We will in the following section see how this picture enables one to achieve the main result of Ref.~\cite{jiang19} quite straightforwardly..

\section{Calculation}

Consider a chiral medium  between the plates, both assumed dielectric.  For simplicity we will neglect the influence from a refractive index, and assume that the vacuum itself has chiral properties. For this situation both the TM and TE modes will rotate between the plates. By that the TM mode partially has turned into a TE mode and vice versa during the propagation between the plates. With angle of rotation $\theta$ the transformation of vector components follows from the matrix of rotation
\begin{eqnarray}
A&=&\left(
\begin{array}{ccc}
\cos\theta & \sin\theta \\
-\sin\theta & \cos\theta\\
\end{array}
\right).
\label{10}
\end{eqnarray}
By the return travel between the plates there will be another angle of rotation $\pm\theta$. With an additional angle $+\theta$ the situation, as mentioned,  is the one of the Faraday effect where  the optical activity is due to a static magnetic field pointing along the direction of propagation. In this case the resulting matrix of rotation for the round trip will be
\begin{eqnarray}
A^2&=&\left(
\begin{array}{ccc}
\cos(2\theta) & \sin(2\theta) \\
-\sin(2\theta) & \cos(2\theta)\\
\end{array}
\right).
\label{11}
\end{eqnarray}

With return rotation $-\theta$ \tc{red}{one} has an optical  medium in zero magnetic field. In this case the round trip rotation will be zero since $A^{-1}A$ is the unit matrix by which the Casimir force is not influenced. In Ref.~\cite{jiang19} the difference between {these} cases does not seem to have been noticed although {this} difference with respect to time reversal symmetry was pointed to.

To obtain the Casimir free energy, from which the Casimir force with a chiral  medium between ideal metal plates can be obtained, we can simply start with the well known Lifshitz result for  metal plates. In the notation of Ref.~\cite{jiang19} this free energy per unit area is
\begin{equation}
E_c=\frac{1}{2\beta}\sum\limits_{n=-\infty}^\infty \int\frac{d^2 k_{||}}{(2\pi)^2}2\ln(1-e^{-2\kappa l}).
\label{12}
\end{equation}
Here $\beta=1/(k_B T)$ where $k_B$ is Boltzmann's constant and $T$ is temperature. The $l$ is the separation between the plates and ${\bf k}_{||}$ is the transverse wave vector along the plates. Further
\begin{equation}
\kappa=\sqrt{\zeta^2+k_{||}^2}
\label{13}
\end{equation}
where $\zeta$ are the Matsubara frequencies that here are normalized  to the values
\begin{equation}
\zeta=\zeta_n=\frac{2\pi n}{\hbar c\beta},
\label{14}
\end{equation}
where $c$ is velocity of light and $n$ is integer. The $\zeta$ is also imaginary frequency, and with normalization (\ref{14})
\begin{equation}
\zeta=\pm i\frac{\omega}{c}
\label{15}
\end{equation}
where either sign may be utilized. Anyway, in the present case with ideal metal surfaces the reflection coefficients do not depend upon frequency by which there will be no further need {for}  Eq.~(\ref{15}) for the situation studied.

With result (\ref{12}) for metal plates the result with a chiral medium between the plates can be obtained by a straightforward extension of it. Only its integrand
\begin{equation}
L=2\ln(1-e^{-2\kappa l})
\label{16}
\end{equation}
is modified. Here the factor 2 {represents} the TM and TE modes (for each set of values for $\zeta$ \tc{red}{and ${\bf k}_{||}$)} that give independent and equal contributions to the free energy. Thus a TM oscillator in one half-plane interacts with a similar TM oscillator in the other half-plane, and likewise with the TE oscillators. Each pair of oscillators {interacts back and forth via the $(e^{-\kappa l})^2=e^{-2\kappa l}$ term.}

Expression (\ref{16}) can be written in matrix notation where the trace is taken
\begin{equation}
L=\rm{Tr}[\ln(I-e^{-2\kappa l}I)]
\label{17}
\end{equation}
with $I$ the unit matrix
\begin{eqnarray}
I&=&\left(
\begin{array}{cc}
1 & 0 \\
0 & 1\\
\end{array}
\right).
\label{18}
\end{eqnarray}

The result with a chiral medium now follows in straightforward way by inserting the rotation matrix (\ref{11}) in expression (\ref{17}). As mentioned earlier the chiral medium will mix the TM and TE modes due to the rotation of the plane of polarization. These modes are like oscillators that oscillate along the $x$ and $y$ directions respectively in both half-planes. With a chiral  medium a coupling will be established between these directions. Its influence will be like a change of vector components by rotation of the coordinate system. Thus for a chiral medium expression (\ref{17}) is modified into
\begin{equation}
L=\rm{Tr}[\ln(I-e^{-2\kappa l}A^2)]=\ln[\rm{det}(I-e^{-2\kappa l}A^2)]
\label{19}
\end{equation}
with $A^2$ given by Eq.~(\ref{11}). The determinant is easily evaluated to obtain
\begin{equation}
\rm{det}(I-e^{-2\kappa l}A^2)=(1-e^{-2\kappa l}\cos(2\theta))^2+(e^{-2\kappa l}\sin(2\theta))^2=1+e^{-4\kappa l}-2e^{-2\kappa l}\cos(2\theta).
\label{20}
\end{equation}
With this inserted in Eq.~(\ref{19}) which further via Eq.~(\ref{17}) is used to replace the corresponding term in Eq.~(\ref{12}), result (C1) of Ref.~\cite{jiang19} is obtained. For temperature $T=0$ the sum in Eq (\ref{12}) can be replaced by integration, and with relation (\ref{14}) one has
\begin{equation}
\frac{1}{2\beta}\sum\limits_{n=-\infty}^\infty \rightarrow\frac{\hbar c}{4\pi}\int\limits_{-\infty}^\infty\,d\zeta=\frac{\hbar c}{2\pi}\int\limits_{0}^\infty\,d\zeta
\label{21}
\end{equation}
by which Eq.~(11) of Ref.~\cite{jiang19} is also recovered where units $\hbar=c=1$ are used.

Note that with $\theta=90\,^o$ the repulsive situation with a magnetic ("metal") plate is recovered with $L=2\ln(1+e^{-2\kappa l})$. This is the Casimir force (66) in Ref.~\cite{hoye18} (with its $A_n=B_n=-1$).


\section{Conclusion}

We  made use of quantum statistical methods developed in  {Refs.~\cite{hoye18,hoye81,hoye92,hoye13,hoye93,brevik88,hoye98}} and elsewhere, in order to treat the Casimir force between ideal metal plates in the presence of an intervening chiral medium. The calculation was made at  finite temperature.  In this way  it was possible to derive the main result of Ref.~\cite{jiang19} (their force expression  Eq.~(C1)), in a very compact manner.

\section*{Acknowledgment}
We acknowledge financial support from the Research Council of Norway, Project 250346.

\end{document}